\documentclass[12pt]{article}
\pdfoutput=1

\usepackage{graphics,amssymb,float}
\usepackage{graphicx}% Include figure files
\usepackage{caption}
\usepackage{rotating}% Include figure files
\usepackage{dcolumn}% Align table columns on decimal point
\usepackage{bm}% bold math
\usepackage{cite}
\usepackage{amsmath}
\usepackage{indentfirst} % activate indent in the first paragraph
\usepackage{epstopdf}
\usepackage[usenames,dvipsnames]{xcolor}
\usepackage{setspace}
\usepackage{hhline}

\usepackage{xcolor}
\usepackage{color}
\usepackage{colortbl}

\makeatletter

\@addtoreset{equation}{section}
\makeatother

\def\lsim{\;\raise0.3ex\hbox{$<$\kern-0.75em\raise-1.1ex\hbox{$\sim$}}\;}
\def\gsim{\;\raise0.3ex\hbox{$>$\kern-0.75em\raise-1.1ex\hbox{$\sim$}}\;}
\def\beq{\begin{equation}}   \def\eeq{\end{equation}}
\def\ba{\begin{array}}       \def\ea{\end{array}}
\def\bea{\begin{eqnarray}}   \def\eea{\end{eqnarray}}

\begin{document}

\begin{titlepage}
%\begin{flushright} LUPM 22-004\\
%\end{flushright}

%\centerline{\bf \today}

\begin{center}
\vspace{1cm}

{\Large\bf Additional Higgs Bosons near 95 and 650 GeV in the NMSSM}

\vspace{2cm}

{\bf{Ulrich Ellwanger$^a$ and Cyril Hugonie$^b$}}\\
\vspace{1cm}
\it $^a$ IJCLab,  CNRS/IN2P3, University  Paris-Saclay, 91405  Orsay,  France\\
ulrich.ellwanger@ijclab.in2p3.fr\\
\it $^b$ LUPM, UMR 5299, CNRS/IN2P3, Universit\'e de Montpellier, 34095 Montpellier, France\\
cyril.hugonie@umontpellier.fr

\end{center}
\vspace{2cm}

\begin{abstract}
Hints for an additional Higgs boson with a mass of about 95 GeV originate from LEP and searches in the diphoton channel by CMS and ATLAS. 
A search for resonant production of SM plus BSM Higgs bosons in the diphoton plus $b\bar{b}$ channel by CMS showed some excess for a 650~GeV resonance decaying into the SM Higgs plus a 95~GeV Higgs boson. We investigate whether these phenomena can be interpreted simultaneously within the NMSSM subject to the latest constraints on couplings of the SM Higgs boson, on extra Higgs bosons from the LHC, and on dark matter direct detection cross sections.
We find that the hints for a 95~GeV Higgs boson in the diphoton channel by CMS and ATLAS and in the diphoton plus $b\bar{b}$ channel by CMS can be fitted simultaneously within the $2\,\sigma$~level.

\end{abstract}

\end{titlepage}

\section{Introduction}

Various well-motivated extensions of the Standard Model (SM) predict additional Higgs bosons, and the search for them is one of the tasks of earlier, present and future experiments in particle physics. These have provided some hints at where such additional Higgs bosons may exist.

The combination of searches for the SM Higgs boson at the ALEPH, DELPHI, L3 and OPAL experiments at LEP \cite{LEPWorkingGroupforHiggsbosonsearches:2003ing} showed some mild excess of events in the $Z^*\to Z+b\bar{b}$ channel in the mass region of $95-100$~GeV. 

Searches for Beyond-the-Standard Model (BSM) Higgs bosons at the LHC in the diphoton channel were performed by CMS and ATLAS. A search at run~1 by CMS showed a $\sim 2\, \sigma$ excess at 97~GeV \cite{CMS-PAS-HIG-14-037}, which was confirmed by CMS later in \cite{CMS:2018cyk} and in \cite{CMS-PAS-HIG-20-002} for a mass hypothesis of 95.4~GeV. A somewhat less sensitive search in the diphoton channel by ATLAS in \cite{ATLAS-CONF-2018-025} lead to an upper limit on the fiducial cross section which did not contradict the possible excess observed by CMS, a recent analysis by ATLAS in the diphoton channel in \cite{ATLAS-CONF-2023-035} showed a mild excess of $1.7\, \sigma$ at 95~GeV.
A search for BSM Higgs bosons in the di-tau channel by CMS in \cite{CMS:2022goy} showed an excess of $2.6\, \sigma$ (local) for a mass of $ 95-100$~GeV.
Finally a search for resonant production via a heavy boson $X$ of a SM Higgs boson together with a BSM Higgs boson $Y$ in the diphoton plus $b\bar{b}$ channel by CMS in \cite{CMS:2023boe} showed an excess of $3.8\,\sigma$ (local) for $M_X\sim 650$~GeV, $M_Y\sim 90-100$~GeV.

The hints for an additional Higgs boson in the mass range of $95-98$~GeV have already lead to numerous explanations within Two-Higgs-Doublet models (2HDM), 2HDMs extended by singlets, radions, pseudo-Goldstone bosons, the Next-to-Minimal supersymmetric extension of the SM (NMSSM) and the $\mu\nu${SSM} \cite{Belanger:2012tt,Cao:2016uwt,Fox:2017uwr,Richard:2017kot,Haisch:2017gql,Biekotter:2017xmf,Liu:2018xsw,Domingo:2018uim,Hollik:2018yek,Wang:2018vxp,Biekotter:2019kde,Cline:2019okt,Choi:2019yrv,Kundu:2019nqo,Sachdeva:2019hvk,Cao:2019ofo,Aguilar-Saavedra:2020wrj,Hollik:2020plc,Abdelalim:2020xfk,Biekotter:2021ovi,Biekotter:2021qbc,Heinemeyer:2021msz,Biekotter:2022jyr,Biekotter:2022abc,Benbrik:2022azi,Benbrik:2022dja,Li:2022etb,Coloretti:2023wng,Banik:2023ecr,Biekotter:2023jld,Azevedo:2023zkg,Escribano:2023hxj,Biekotter:2023oen,Belyaev:2023xnv,Aguilar-Saavedra:2023vpd,Dutta:2023cig,Maniatis:2023aww,Cao:2023gkc}.

The $\sim 2\, \sigma$ excess at LEP was quantified in \cite{Cao:2016uwt}. Let us denote the extra (lighter) Higgs boson by $H_1$, with a reduced coupling to vector bosons $W^\pm, Z$ (relative to the coupling of a SM-like Higgs boson of corresponding mass) given by $C_V(1)$. Then the authors in \cite{Cao:2016uwt} define (see also \cite{Biekotter:2022jyr})
\beq\label{mulep}
\mu^{LEP}_{bb} \equiv C_V(1)^2\times BR(H_1\to b\bar{b})/BR(H_{SM}^{95}\to b\bar{b})= 0.117 \pm 0.057
\eeq
where $H_{SM}^{95}$ denotes a SM-like Higgs boson with a mass of 95~GeV.

The best fits for a diphoton signal of $H_1$ in CMS and ATLAS were combined in \cite{Biekotter:2023oen}. The authors in \cite{Biekotter:2023oen} obtain
\beq\label{mugamgam}
\mu_{\gamma\gamma}^{LHC} = \frac{\sigma(gg \to H_1\to \gamma\gamma)}{\sigma(gg \to H_{SM}^{95} \to \gamma\gamma)}
=  {0.24^{+0.09}_{-0.08}}\; .
\eeq
Again, $H_{SM}^{95}$ denotes a SM-like Higgs boson with a mass of 95~GeV.

The best fit for the excess in the di-tau channel at 95~GeV observed by CMS in \cite{CMS:2022goy} corresponds to a cross section times branching fraction
\beq\label{ditau}
\sigma(gg \to H_1 \to \tau\tau)= 7.8^{+3.9}_{-3.1}\text{pb}\; .   %\; ,\qquad
\eeq
For $H_{SM}^{95}$ we obtain
\beq\label{ditau1}
\mu_{\tau\tau}^{LHC} = \frac{\sigma(gg \to H_1\to \tau\tau)}{\sigma(gg \to H_{SM}^{95} \to \tau\tau)}
= 1.38^{+0.69}_{-0.55}\; .
\eeq

Finally the best fit for the excess in the search for $X \to (H_{SM}\to \gamma\gamma) + (H_1 \to b\bar{b})$ for $M_X\simeq 650$~GeV and $M_{H_1}=90-100$~GeV observed by CMS in \cite{CMS:2023boe} is a cross section times branching fraction given by
\beq\label{650GeV}
\sigma_{bb\gamma\gamma}=\sigma(gg \to X_{650} \to (H_1 \to b\bar{b}) + (H_{SM}\to \gamma\gamma) ) = 0.35^{+0.17}_{-0.13}\,\text{fb}\; .
\eeq
In fact, a search for $X \to (H_1 \to b\bar{b})+(H_{SM}\to \tau\tau) $ 
has been carried out by CMS in \cite{CMS:2021yci}, without an excess for $M_{H_1}=90-100$~GeV and $M_X= 600$~GeV or $M_X= 700$~GeV. Instead, an upper 95\% CL limit of $\sim 3$\, fb was obtained for
{$\sigma_{bb\tau\tau}$} for these choices of masses. For $H_{SM}$, the $BR(H_{SM}\to \tau\tau)$ is about 30 times larger than the $BR(H_{SM}\to \gamma\gamma)$. Accordingly, assuming $3$\, fb as upper limit on $X_{650} \to (H_{SM}\to \tau\tau) + (H_1 \to b\bar{b})$ implies an upper 95\% CL limit of $\sim 0.1$\,fb on $X_{650} \to (H_{SM}\to \gamma\gamma) + (H_1 \to b\bar{b})$ which is barely (but still) compatible with the lower $2\,\sigma$ boundary of $0.09$\,fb of the fit in eq.\eqref{650GeV}.
(The 650~GeV excess in \cite{CMS:2023boe} had already been discussed in connection with the 95~GeV excesses in \cite{Banik:2023ecr} and \cite{Azevedo:2023zkg}.)

The aim of the present paper is to verify in how far the previous excesses can be described simultaneously within the NMSSM \cite{Maniatis:2009re,Ellwanger:2009dp} subject to the most recent constraints from the LHC, notably the recent measurements of Higgs couplings by CMS \cite{CMS:2022dwd} and ATLAS \cite{ATLAS:2022vkf}, the upper limit on the dark matter relic density (allowing for additional contributions beyond the lightest supersymmetric particle) and searches for direct detection of dark matter \cite{CRESST:2015txj,DarkSide:2018bpj,XENON:2018voc,XENON:2019rxp,PICO:2019vsc,PandaX-4T:2021bab,LZ:2022lsv}. To this end we employ the public codes \texttt{NMSSMTools-6.0.2} \cite{Ellwanger:2004xm,Ellwanger:2005dv,NMSSMTools} and MicrOMEGAs \cite{Belanger:2013oya}.
Similar studies of excesses within the NMSSM have been performed before in \cite{Belanger:2012tt,Cao:2016uwt,Biekotter:2017xmf,Domingo:2018uim,Wang:2018vxp,Cao:2019ofo,Choi:2019yrv,Hollik:2020plc,Biekotter:2021qbc,Li:2022etb} without, however, the most recent constraints from the LHC and, notably, without considering the possible excess in $X_{650} \to (H_{SM}\to \gamma\gamma) + (H_1 \to b\bar{b})$.
{Given the above constraints, we find that the hints for a 95~GeV Higgs boson at LEP and in the diphoton channel by CMS and ATLAS and in the diphoton plus $b\bar{b}$ channel by CMS can be fitted simultaneously within the $2\,\sigma$~level.}

In the next Section we summarize the relevant features of the NMSSM, and the constraints which we apply to our scan of the parameter space of the NMSSM. In Section~3
we show the results of scans of the NMSSM parameter space in the form of figures showing correlations among masses and production cross sections relevant for searches for additional heavy resonances in various channels. We conclude in Section~4.

\section{Applied constraints to the NMSSM}

The Higgs sector of the NMSSM consists in two SU(2) doublets and a complex SU(2) singlet. In the CP-conserving NMSSM, the physical scalars can be decomposed into three neutral CP-even states, two neutral CP-odd states and one complex charged state. One of the three neutral CP-even states has to correspond to the SM-like Higgs boson. A priori the masses and couplings of the remaining states can assume a large range of values, depending on the five NMSSM-specific parameters $\lambda$, $\kappa$, $A_\lambda$, $A_\kappa$, $\mu_{\rm eff}$ as well as on $\tan\beta$ \cite{Maniatis:2009re,Ellwanger:2009dp}. 

In general, the three neutral CP-even states as well as the two neutral CP-odd states are mixtures of SU(2) doublets and a SU(2) singlet; thereby all scalars obtain couplings to SM fermions and gauge bosons (originally reserved to the SU(2) doublets). Still, in most of the parameter space one can denote each of the three CP-even scalars $H_1$, $H_2$ and $H_3$ (ordered in mass) as either mostly singlet-like, or mostly SM-like, or mostly MSSM-like. (Pure singlet-like, SM-like or MSSM-like states represent the so-called Higgs basis.)

The mostly singlet-like state is a candidate for an extra BSM Higgs boson $H_1$ near 95~GeV \cite{Belanger:2012tt,Cao:2016uwt,Biekotter:2017xmf,Domingo:2018uim,Wang:2018vxp,Cao:2019ofo,Choi:2019yrv,Hollik:2020plc,Biekotter:2021qbc,Li:2022etb}. However, as discussed below, the recent combinations of CMS \cite{CMS:2022dwd} and ATLAS \cite{ATLAS:2022vkf} of the couplings of the SM Higgs boson in the $\kappa$ framework limit the couplings of the singlet-like state. Consequently, its remaining allowed production cross sections at LEP and the LHC contradict some of the scenarios proposed earlier.

The notion MSSM-like refers to a nearly degenerate SU(2) doublet (if much heavier than the SM-like Higgs boson) consisting in a neutral CP-even, a neutral CP-odd and a charged complex state. The CP-even state $H_3$ is a candidate for a heavy resonance $X$ near $650$~GeV generating the excess in $X \to (H_{SM}\to \gamma\gamma) + (H_1 \to b\bar{b})$ observed by CMS \cite{CMS:2023boe}. Expressions for triple Higgs couplings in the NMSSM have been given in \cite{Ellwanger:2022jtd}; for the triple Higgs coupling relevant here (recall that $H_{SM}=H_2$) one finds at tree level
\beq\label{H1H2H3}
-H_1 H_2 H_3\left(\sqrt{2}\kappa \mu_{\rm eff} +\frac{\lambda}{\sqrt{2}} A_\lambda\right) + \dots\; ,
\eeq
where the dots denote relatively small corrections originating from the rotation from the Higgs basis to the physical basis.
The production of $X=H_3$ at the LHC can well take place via gluon fusion. We recall, however, that the production cross section times branching fraction for $X \to (H_{SM}\to \gamma\gamma) + (H_1 \to b\bar{b})$ is limited to $\sim 0.1$\, fb by upper limits on $X \to (H_{SM}\to \tau\tau) + (H_1 \to b\bar{b})$ from CMS in \cite{CMS:2021yci}.

The scan of the general NMSSM parameter space is performed with help of the codes \texttt{NMSSMTools-6.0.2} \cite{Ellwanger:2004xm,Ellwanger:2005dv,NMSSMTools} and MicrOMEGAs \cite{Belanger:2013oya}. 

{
For the SM-like Higgs boson we require a mass within $125.2\pm 3$~GeV (allowing for theoretical uncertainties), and that the couplings in the $\kappa$-framework satisfy combined limits of CMS \cite{CMS:2022dwd} and ATLAS \cite{ATLAS:2022vkf}. In the NMSSM, the reduced couplings of $H_{SM}$ to $W$ and $Z$ bosons are the same, whereas they are measured separately by ATLAS and CMS. However, since the corresponding uncertainties are correlated, one cannot consider these measurements as independent. Given that the measurements of $\kappa_Z$ are slightly more precise, we combine the corresponding results of ATLAS and CMS and ignore the measurements of $\kappa_W$ in order to remain conservative. From Fig.~6 in \cite{ATLAS:2022vkf} we find including $1\,\sigma$ uncertainties $\kappa_Z = 0.99\pm 0.057$, from Fig.~4a in \cite{CMS:2022dwd} we use $\kappa_Z = 1.04\pm 0.07$. Combining both measurements one obtains $\kappa_{Z} > 0.923$ at the $2\,\sigma$ level. {This value of $\kappa_{Z}$ close to 1.00 corresponds to the so-called alignment limit of the NMSSM discussed in \cite{Carena:2015moc}, and the values of $\lambda$ and $\tan\beta$ are indeed in the range found in \cite{Carena:2015moc}. It has been proposed previously in \cite{Biekotter:2021qbc} that the alignment limit of the NMSSM can accomodate an extra Higgs boson near 95~GeV.} In BSM models with an arbitrary number of Higgs doublets and singlets one obtains the sum rule $\sum_i C_V(i)^2 = 1$, hence 
\beq
C_V(1)^2 < 0.148
\eeq
at the $2\,\sigma$ level.
}

In addition we impose constraints from b-physics, constraints from searches for BSM Higgs bosons by ATLAS and CMS as implemented in \texttt{NMSSMTools-6.0.2}, and constraints from the absence of a Landau singularity for the Yukawa couplings below the GUT scale. It confines values of the NMSSM-specific coupling $\lambda$ to $\lambda \lsim 0.7$. Constraints from the anomalous magnetic moment of the muon as in \cite{Domingo:2022pde} are left aside as these concern the smuon/gaugino sector which is irrelevant here. (Constraints from the anomalous magnetic moment of the muon can always be satisfied by choosing the soft supersymmetry breaking trilinear coupling $A_\mu$ large enough.)
The constraint on $M_W$ as applied in \cite{Domingo:2022pde} is not used since it relies on a single experimental result which differs significantly from many others. 
{The references to constraints from BSM Higgs-boson searches, b-physics (of little relevance here) are listed on the web page {\sf https://www.lupm.in2p3.fr/users/nmssm/history.html}.}
All soft supersymmetry breaking terms are taken below 3~TeV.
Constraints on the sparticle spectrum are taken into account using the code \texttt{SModels-2.2.0} \cite{Kraml:2013mwa,Dutta:2018ioj,Khosa:2020zar,Alguero:2021dig}.

We require that the lightest supersymmetric particle (LSP) is neutral (the lightest neutralino), since it is stable and contributes necessarily to the relic density of the universe. We do not require that it accounts for {\it all} of the observed dark matter relic density as there may exist additional contributions from physics far above the weak scale.
However, the stable lightest neutralino unavoidably contributes to dark matter direct detection experiments, and must satisfy corresponding constraints which are imposed since the properties of the lightest neutralino (its mass and its annihilation rate typically via the CP-odd scalar $A_1$ in the s-channel) depend on parameters which play also a role in the NMSSM Higgs sector. We find that the LSP is a higgsino-singlino mixture, with a relic density {$\Omega h^2 \approx 10^{-4}-10^{-3}$}.

For the calculation of the cross sections $ggF\to H/A$ we start with the BSM Higgs production cross sections at $\sqrt{s} = 13$~TeV from the LHC Higgs Cross Section Working Group \cite{LHCyellowreport} (CERN Yellow Report~4 2016). These are multiplied by the reduced couplings squared of $H/A$. Thereby we capture most of the radiative QCD corrections in the form of K-factors; the remaining theoretical uncertainties are at most of ${\cal O}(10\%) $.

For the purpose of this paper we require that the singlet-like scalar has a mass in the range $95.4\pm 3$~GeV (allowing for a theoretical uncertainty of 3~GeV), $\mu^{LEP}_{bb}$ in the $2\, \sigma$ range of \eqref{mulep},
and $\mu_{\gamma\gamma}^{LHC}$ in the $2\, \sigma$ range of \eqref{mugamgam}. In order to describe the excess in $\sigma_{bb\gamma\gamma}$, we require that the MSSM-like scalar $H_3$ has a mass in the range $650\pm 25$~GeV (given that the mass $M_X$ in \cite{CMS:2023boe} is given in steps of $650\pm n\times 50$~GeV), and $\sigma_{bb\gamma\gamma}$ in the $2\, \sigma$ range of \eqref{650GeV}.

Let us discuss in how far the excesses $\mu^{LEP}_{bb}$ in \eqref{mulep}, $\mu_{\gamma\gamma}^{LHC}$ in \eqref{mugamgam}, $\mu_{\tau\tau}^{LHC}$ in \eqref{ditau1} and $\sigma_{bb\gamma\gamma}$ in \eqref{650GeV} can be described simultaneously.
First, once the contribution of ATLAS to $\mu_{\gamma\gamma}^{LHC}$ from \cite{ATLAS-CONF-2023-035}
is combined with the corresponding contributions from CMS {implying a lower central value} as in \cite{Biekotter:2023oen}, 
{the excesses $\mu^{LEP}_{bb}$ and $\mu_{\gamma\gamma}^{LHC}$ can be described simultaneously at the $2\,\sigma$ level in the NMSSM with its type~II Yukawa structure. Within the $2\,\sigma$ level, a suppression of the $BR(H_1 \to b\bar{b})$ in order to enhance the $BR(H_1 \to \gamma\gamma)$ (as argued earlier in \cite{Biekotter:2022jyr}) is then no longer necessary.}
Also the excess in $\sigma_{bb\gamma\gamma}$ in \eqref{650GeV} can be fitted simultaneously at the $2\,\sigma$ level. 

{
However, a description of the di-tau excess $\mu_{\tau\tau}^{LHC}$ in \eqref{ditau1} would require a large $BR(H_1 \to \tau^+\tau^-)$ (or a large $H_1$ production cross section) which is incompatible with present constraints on the $H_1 - H_{SM}$ mixing angle.
}
The incompatibility of the $\mu_{\gamma\gamma}^{LHC}$ and $\mu_{\tau\tau}^{LHC}$ for a type II Yukawa structure was also underlined in \cite{Biekotter:2023oen}. Thus we will not require a description of the di-tau excess $\mu_{\tau\tau}^{LHC}$ in the following.

{Then, all constraints from eqs.\eqref{mulep}, \eqref{mugamgam} and \eqref{650GeV} can be satisfied simultaneously. However, the upper $2\,\sigma$ limit on $\kappa_\tau$ from the combination of \cite{CMS:2022dwd} and \cite{ATLAS:2022vkf}, 
\beq\label{ktau}
\kappa_\tau < 1.033\, ,
\eeq
constrains the parameters and the cross sections to very narrow ranges around values corresponding to those of the benchmark point BP1 shown in the next Section. }

{
We found it appropriate to show the possible parameters and cross sections after relaxing the constraint \eqref{ktau}.
Then the input parameters assume values within the ranges shown in Table~1. We show $\mu^{LEP}_{bb}$ and $\mu_{\gamma\gamma}^{LHC}$ for viable points as function of $M_{H_3}$ in Fig.~1, and $\sigma_{bb\gamma\gamma}$ and $\sigma_{bb\tau\tau}$ in Fig.~2.
Within these and the subsequent Figures, the light blue regions contain points which satisfy the constraint \eqref{ktau} on $\kappa_\tau$.
}
As discussed in the Introduction, both $\sigma_{bb\gamma\gamma}$ and $\sigma_{bb\tau\tau}$ are limited from above by constraints from the search by CMS in \cite{CMS:2021yci}.
The coloured dots in all figures indicate six benchmark points (BPs) BP1 (red), BP2 (green), BP3 (blue), BP4 (yellow), BP5 (violet) and BP6 (orange), whose properties are given in the Tables~2 and 3 in the next Section.

\begin{table}[h!]
\begin{center}
\begin{tabular}{| c | c | c | c | c | c |  }
\hline
 $\lambda$  & $\kappa$  & $A_\lambda$ &$A_\kappa$ & $\mu_{\rm eff}$ & $\tan\beta$ 
\\
\hline
$0.610 - 0.687$ & $0.307 - 0.391$ & $400 - 480$ & $-621 -(-402)$ & $238 - 291$ & $1.97 - 2.58$   \\
\hline
\hhline{|=|=|=|=|=|=|}
 $M_1$ & $M_2$ &$M_3$ &  $A_t$ & $M_{Q_3}$  & {$M_{U_3}$ }
\\
\hline 
$255-3000$ & $338-2800$ & $423-3000$ & $-2222 - 1288$ & $825 - 3000$ & $857-3000$
 \\
\hline
\end{tabular}
\end{center}
\caption{Range of input parameters for our scan (dimensionful parameters in GeV). }
\label{tab:1}
\end{table}

\begin{figure}[ht!]
\begin{center}
\hspace*{-10mm}
\begin{tabular}{cc}
\includegraphics[scale=0.32]{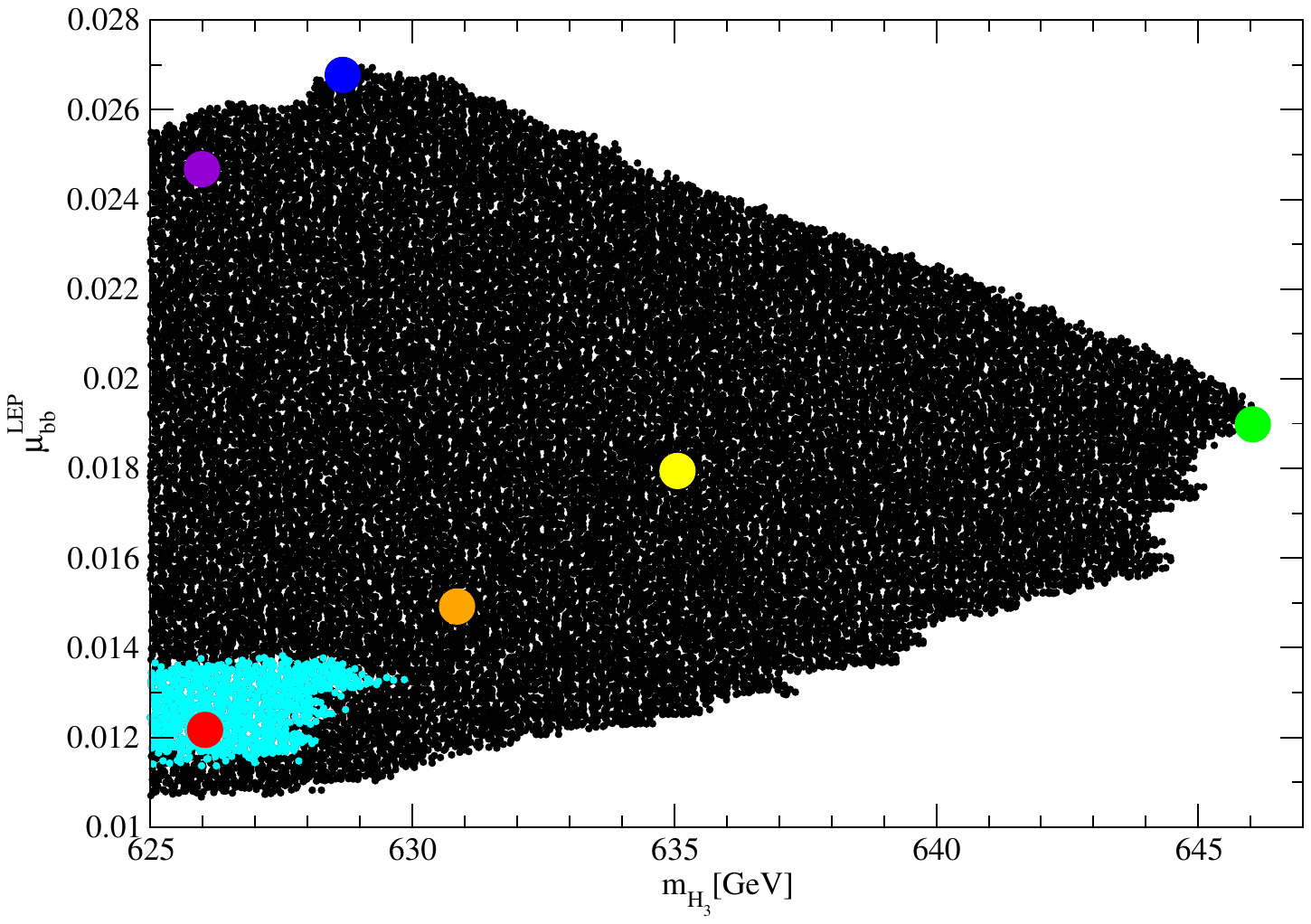}
   & % \hskip -0.5cm
\includegraphics[scale=0.32]{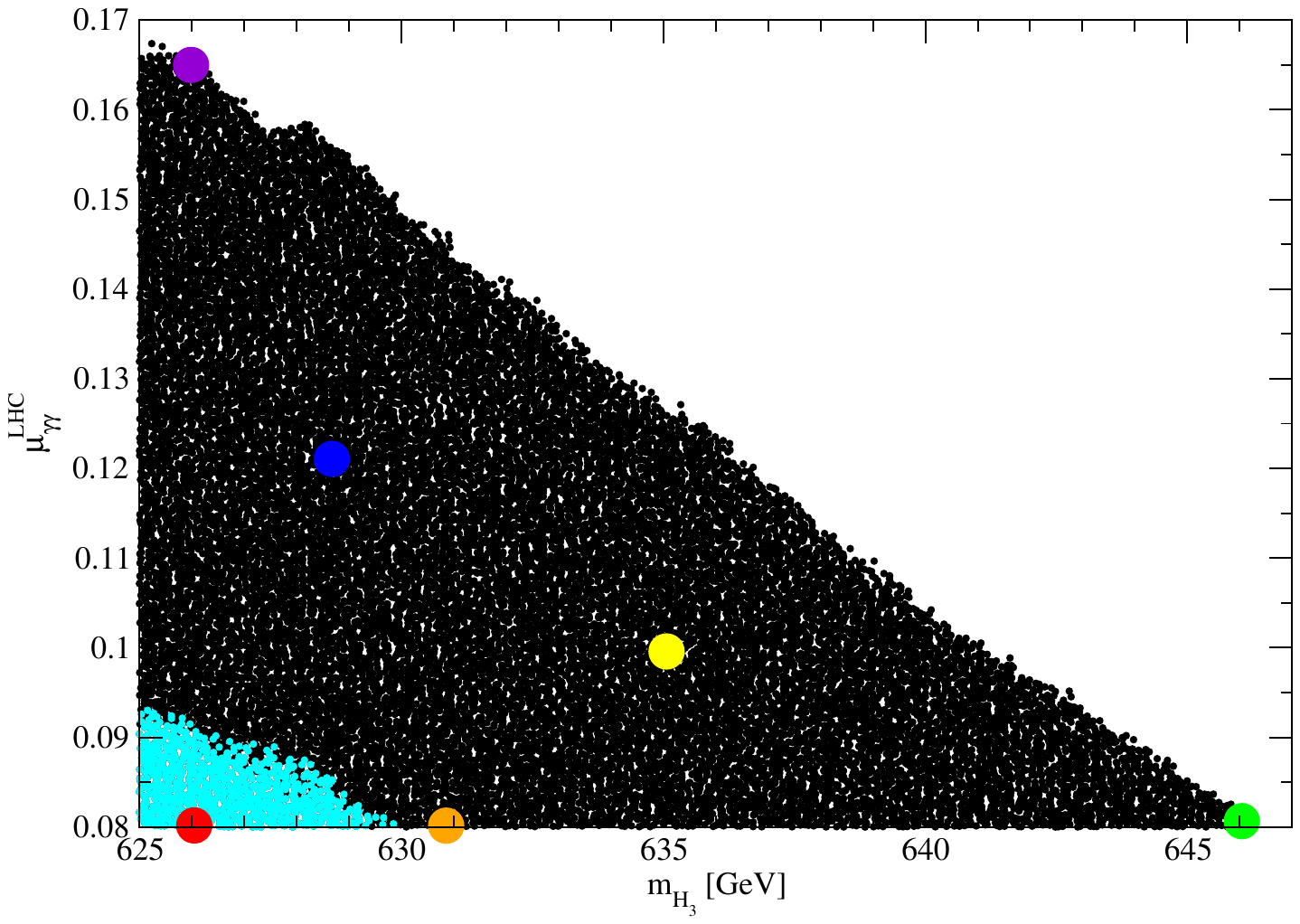}
\end{tabular}
\end{center}
\vskip -1.cm
\caption{$\mu^{LEP}_{bb}$ as function of $M_{H_3}$ (left), $\mu_{\gamma\gamma}^{LHC}$ as function of $M_{H_3}$ (right). The coloured dots here and the subsequent figures denote six benchmark points whose properties are given in the Tables~2 and 3.}
\label{fig:1}
\end{figure}

\begin{figure}[ht!]
\begin{center}
\hspace*{-10mm}
\begin{tabular}{cc}
\includegraphics[scale=0.32]{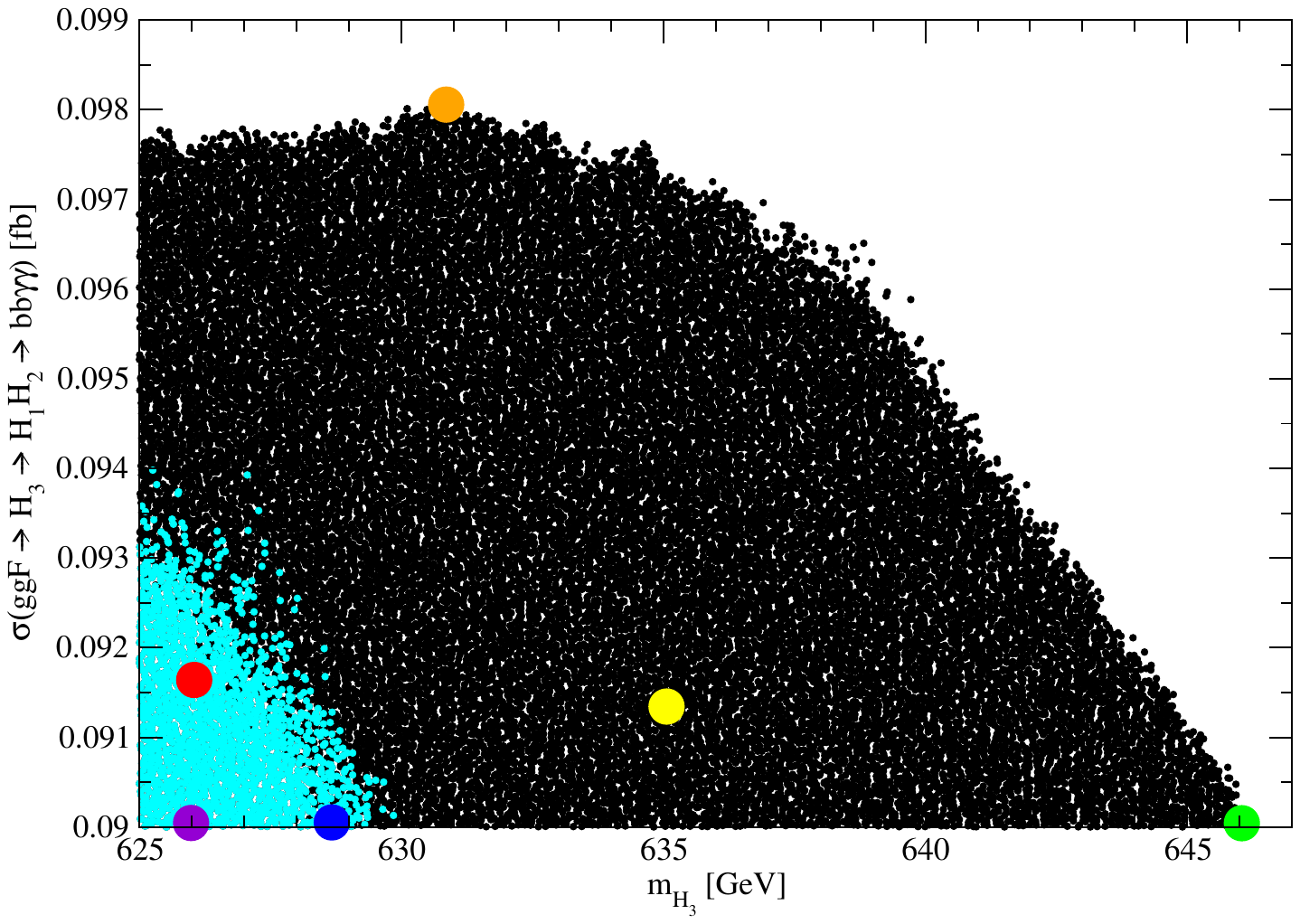}
   & % \hskip -0.5cm
\includegraphics[scale=0.32]{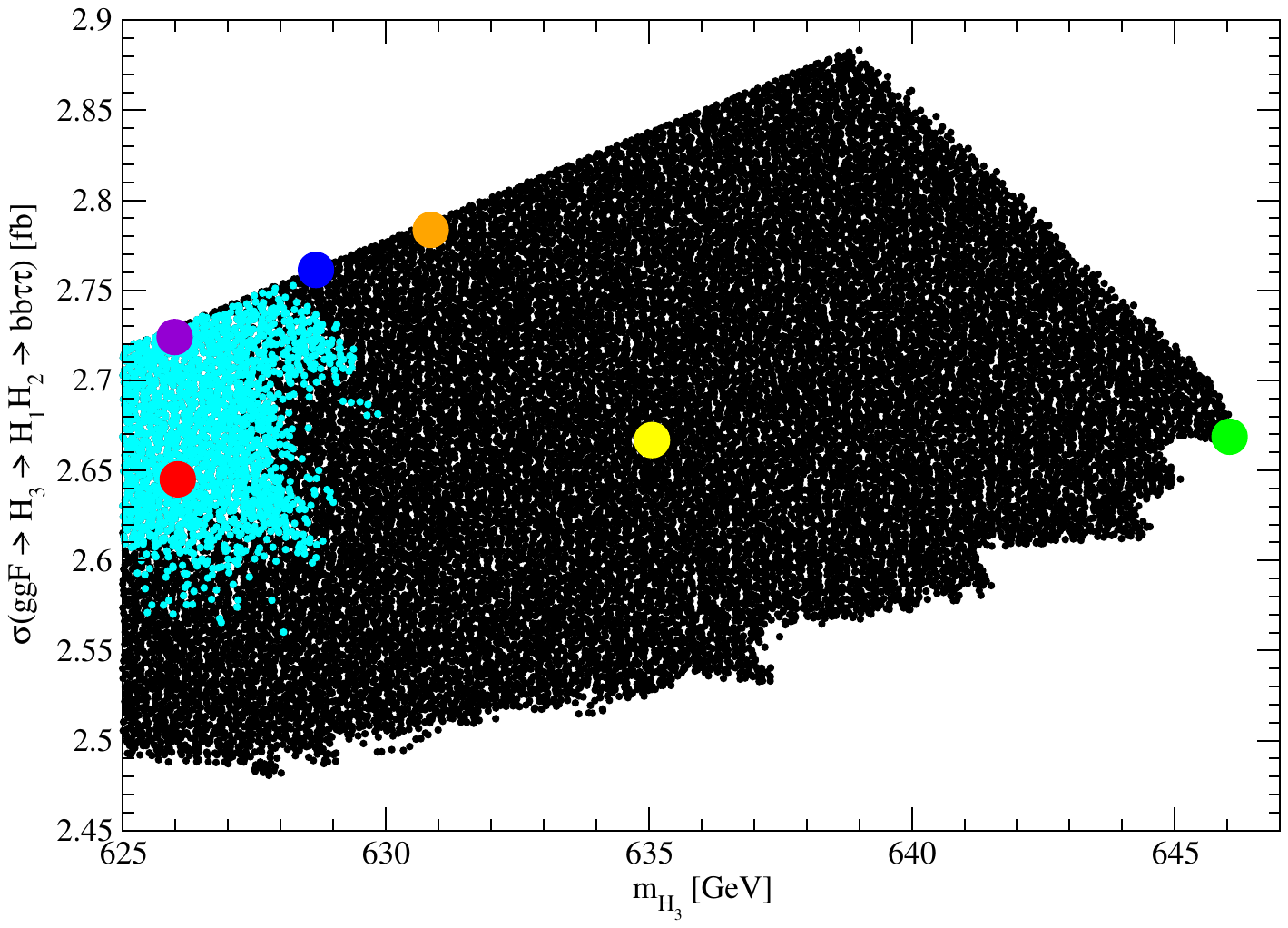}
\end{tabular}
\end{center}
\vskip -1.cm
\caption{$\sigma_{bb\gamma\gamma}$ as function of $M_{H_3}$ (left), $\sigma_{bb\tau\tau}$ as function of $M_{H_3}$ (right). $\sigma_{bb\gamma\gamma}$ and $\sigma_{bb\tau\tau}$ are limited from above by constraints from the search by CMS in \cite{CMS:2021yci}.}
\label{fig:2}
\end{figure}

{As a consequence of the freedom in the elements of the $3\times 3$ mass matrix in the CP-even Higgs sector, the coupling of $H_S$ to b-quarks and thus the branching fraction $BR(H_S\to b\bar{b})$ is variable. Since this branching fraction is dominant, its reduction implies an increase of others like the diphoton rate $BR(H_S\to \gamma\gamma)$. 
%(The diphoton rate could also be enhanced by larger couplings of $H_S$ to top quarks \cite{Biekotter:2021qbc} or chargino loops.) 
At first sight, an increase of the diphoton rate is welcome in order to fit the central value of $\mu_{\gamma\gamma}^{LHC}$ \cite{Biekotter:2021qbc}. However, since the increase of the diphoton rate comes hand-in-hand with a reduced $BR(H_S\to b\bar{b})$, it becomes difficult to obtain a large enough $\sigma_{bb\gamma\gamma}$, which is left aside in \cite{Biekotter:2021qbc}. In fact, since $\sigma_{bb\gamma\gamma}$ decreases with $M_{H_3}$ for kinematic reasons, larger values of $M_{H_3}$ require larger values of the $BR(H_S\to b\bar{b})$ and thus smaller values of the $BR(H_S\to \gamma\gamma)$ or $\mu_{\gamma\gamma}^{LHC}$. This explains the shape of Fig.1 (right).
Actually we find that the desired cross section for $\sigma_{bb\gamma\gamma}$ can be achieved only for {$M_{H_3}\lsim 645$~GeV.}}

\section{Benchmark Planes and Points}

If the scenario with additional Higgs bosons near 95~GeV and 650~GeV is realized within the NMSSM, several additional search channels can serve to discover or to exclude it. In this section we present the prospects for such additional searches in the form of benchmark planes of couplings and cross sections.

A Higgs resonance $H_3$ near 650~GeV can be searched for by its decays into heavy quarks. We find that the region in the NMSSM parameter space satisfying the constraints corresponds to relatively small values of $\tan\beta \sim 2-3$. Then the search for the $H_3$ decay into $b\bar{b}$ is not very promising, in contrast to the search for its decay into $t\bar{t}$ as performed by CMS in \cite{CMS:2019pzc}. In Fig.~3 left we show its coupling strength modifier $g_{Htt}$ as a function of the heavy scalar boson mass. The width of $H_3$ ($\sim 6$~GeV) is always $\sim 1\%$ of its mass which is relevant for the search in this channel. Nearly degenerate with $H_3$ (about 3~GeV lighter) is a pseudoscalar $A_2$, with a width of $\sim 1.5\%$ of its mass and with a very similar coupling strength modifier $g_{Att}$ shown in Fig.~3 right. The upper limits from \cite{CMS:2019pzc} are $\sim .735$ on $g_{H_3tt}$ and $\sim .675$ on $g_{A_2tt}$. It should be noted that the branching fractions of these states into $t\bar{t}$ could be somewhat reduced by up to $\approx 10\%$ due to decays into neutralinos/charginos depending on the parameters of this sector. Still, given that the limits from \cite{CMS:2019pzc} are based on an integrated luminosity at the LHC of $35.9$~fb$^{-1}$, corresponding updates may well become sensitive to the NMSSM scenarios presented here.

\begin{figure}[ht!]
\begin{center}
\hspace*{-10mm}
\begin{tabular}{cc}
\includegraphics[scale=0.32]{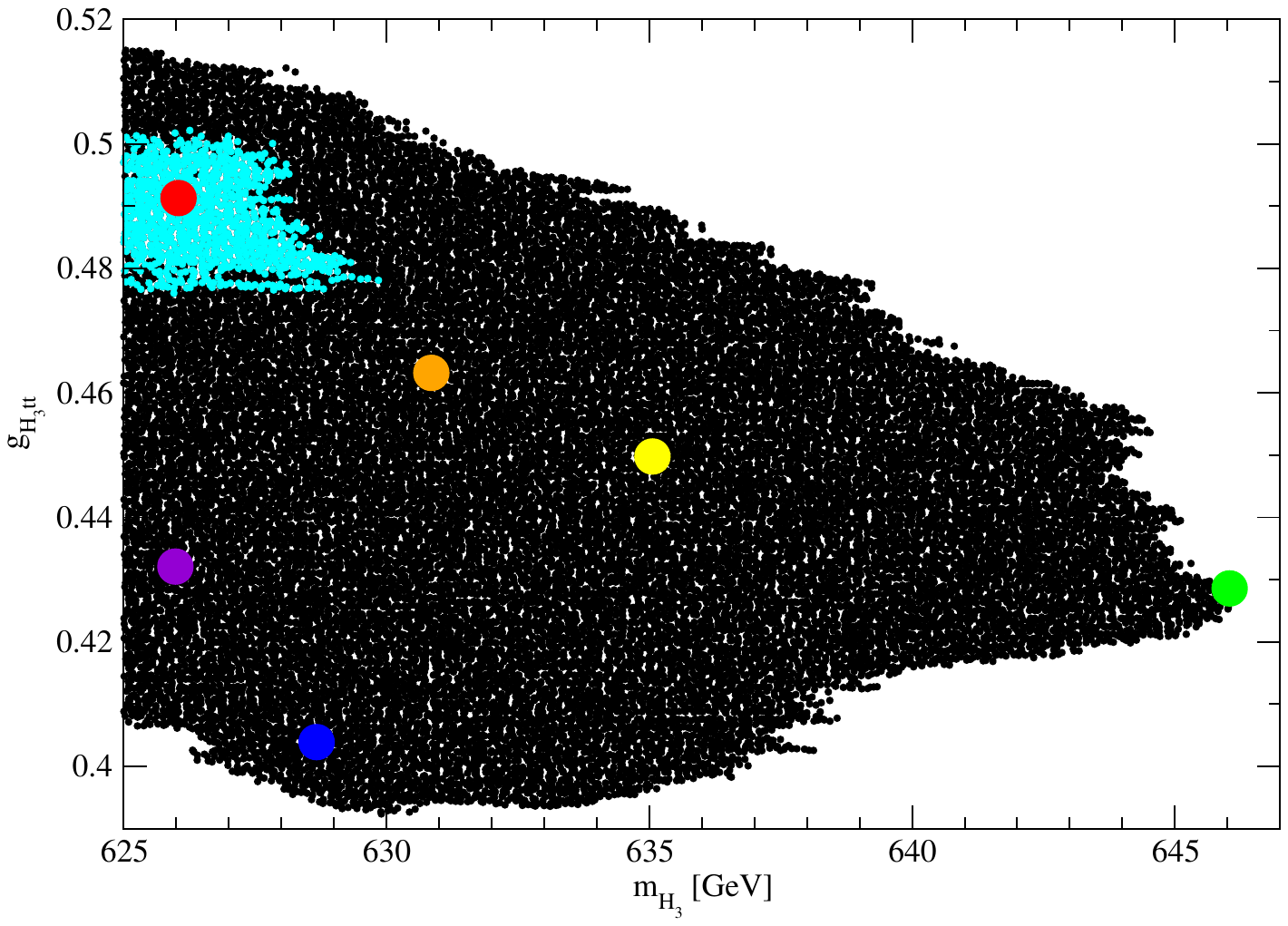}
   & % \hskip -0.5cm
\includegraphics[scale=0.32]{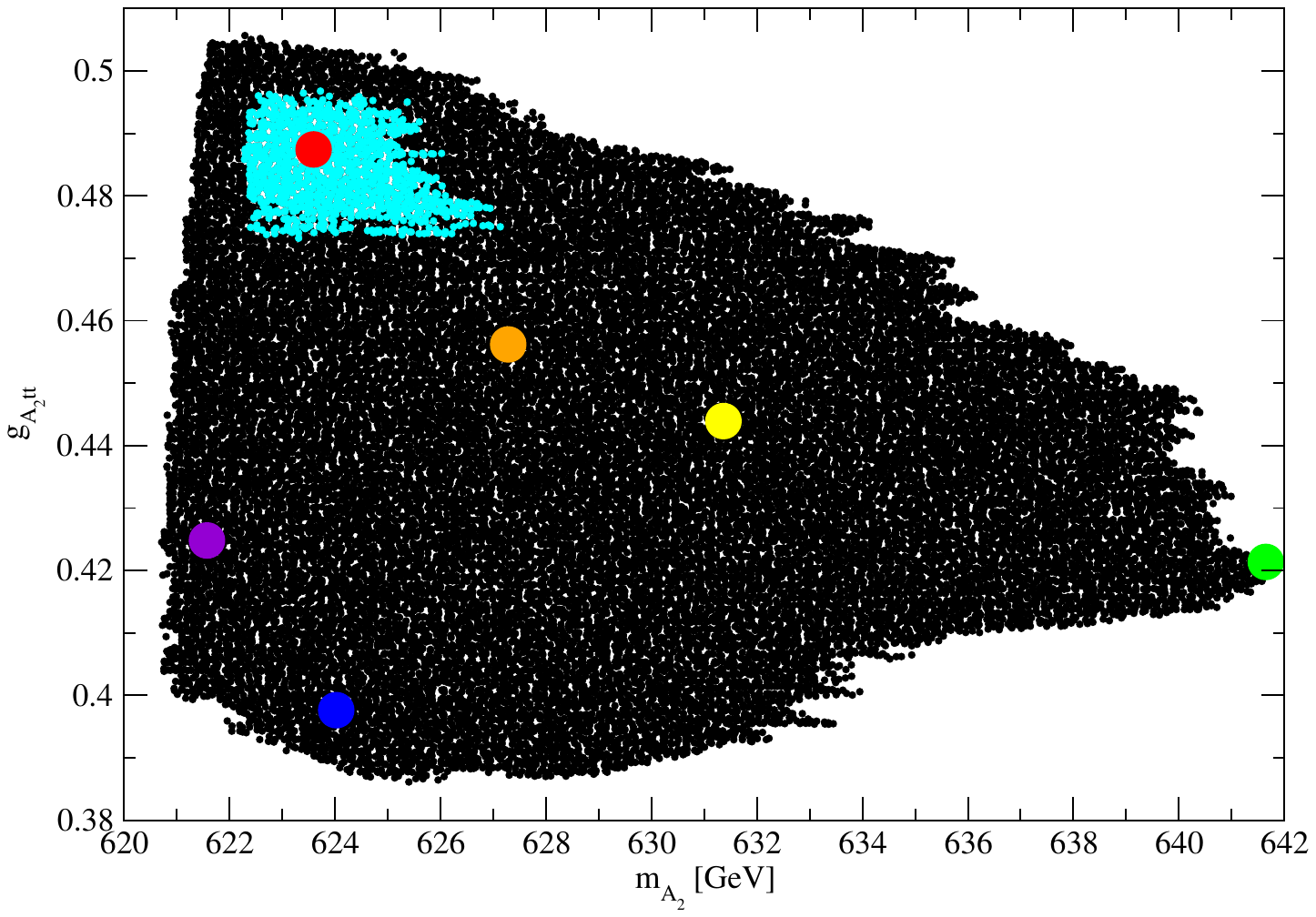}
\end{tabular}
\end{center}
\vskip -1.cm
\caption{Left: Coupling strength modifier $g_{H_3tt}$ as a function of the heavy scalar boson mass $M_{H_3}$. Right: Coupling strength modifier $g_{A_2tt}$ as a function of the heavy pseudo-scalar boson mass $M_{A_2}$. The upper limits from \cite{CMS:2019pzc} are $\sim .735$ on $g_{H_3tt}$ and $\sim .675$ on $g_{A_2tt}$.}
\label{fig:3}
\end{figure}

The coupling $H_2 H_2 H_3$ contributes to resonant SM Higgs pair production on which the most recent constraints originate from ATLAS in \cite{ATLAS-CONF-2021-052,ATLAS:2021ifb} and CMS in \cite{CMS-B2G-20-007,CMS:2021roc}, the latter only for heavy resonances above 800~GeV. From the combination of final states in \cite{ATLAS-CONF-2021-052} and for $M_{H_3}$ near $650$~GeV, the upper limit on $\sigma\times Br(ggF\to H_3\to H_{SM}+H_{SM})$ is $\sim 11$~fb. However, one finds that the coupling $H_2 H_2 H_3$ is suppressed by $M_Z$ and, for the allowed regions of the parameter space of the NMSSM, much smaller than the $H_1 H_2 H_3$ coupling in \eqref{H1H2H3} implying a maximal cross section of $\sim 1$~fb for resonant SM Higgs pair production for $M_{H_3}$ near $650$~GeV. 
Actually a mild $\sim 1\, \sigma$ excess is visible in \cite{ATLAS:2021ifb} for the $b\bar{b}\gamma\gamma$ channel for $M_{H_3}$ near $650$~GeV, but the required cross section for a visible excess in this channel would be impossible to achieve within the allowed regions of the parameter space of the NMSSM.

Relatively large cross sections of $\approx 10$~fb are found within the allowed regions of the parameter space of the NMSSM for the process $ggF\to H_3\to H_1+H_1$, with branching ratios of $H_1$ into $b\bar{b}$, $\tau^+\tau^-$ and $\gamma\gamma$ $\sim 20\%$ larger than for $H_{SM}$.
We find it worthwhile to perform corresponding searches in channels with low enough SM backgrounds; they may lead to hints for or the discovery of two BSM Higgs bosons at a time. Corresponding cross sections times branching fractions are shown in Figs.~4. (No upper limits exist on these processes at present.)

\begin{figure}[ht!]
\begin{center}
\hspace*{-10mm}
\begin{tabular}{cc}
\includegraphics[scale=0.32]{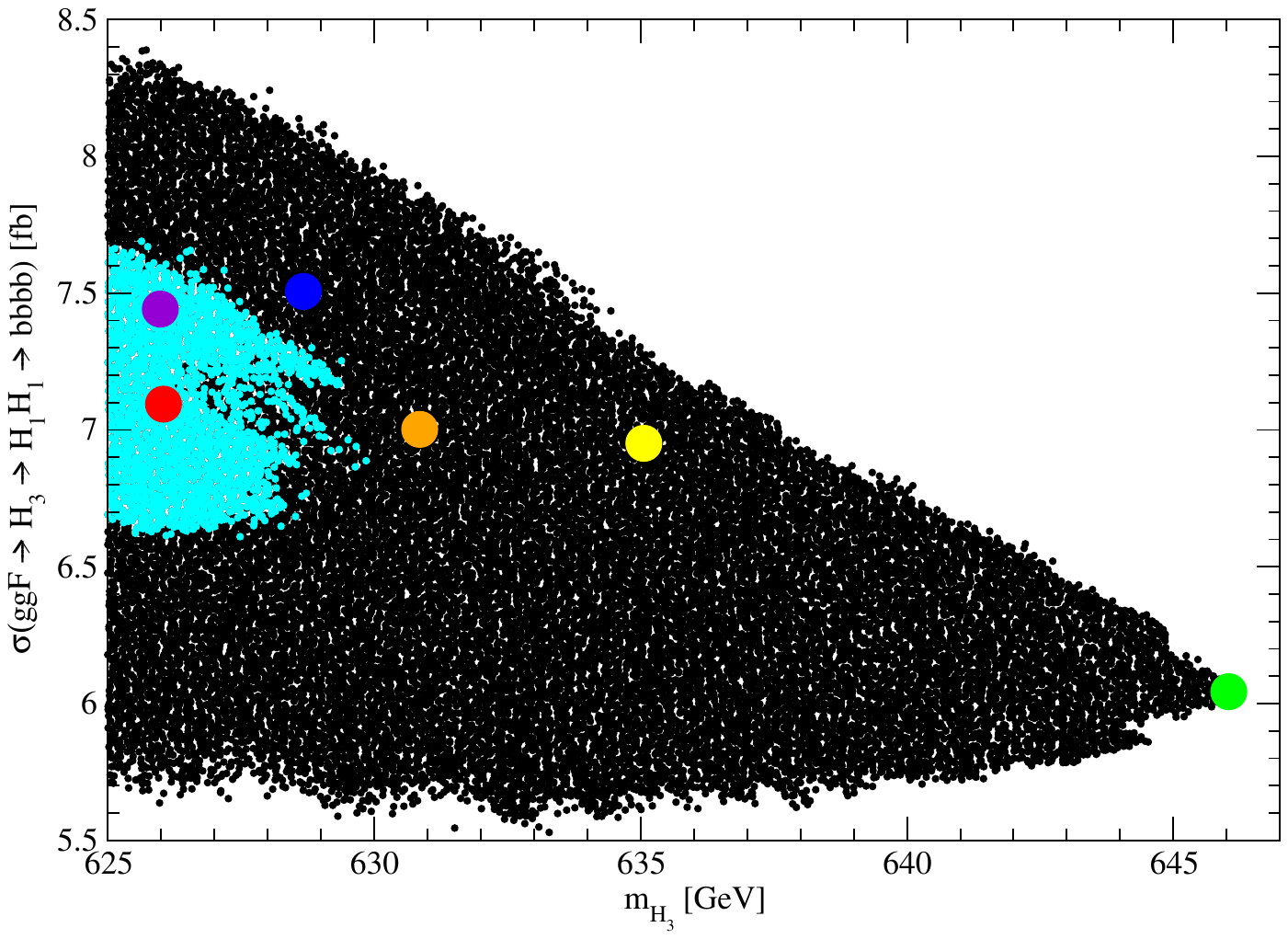}
   & % \hskip -0.5cm
\includegraphics[scale=0.32]{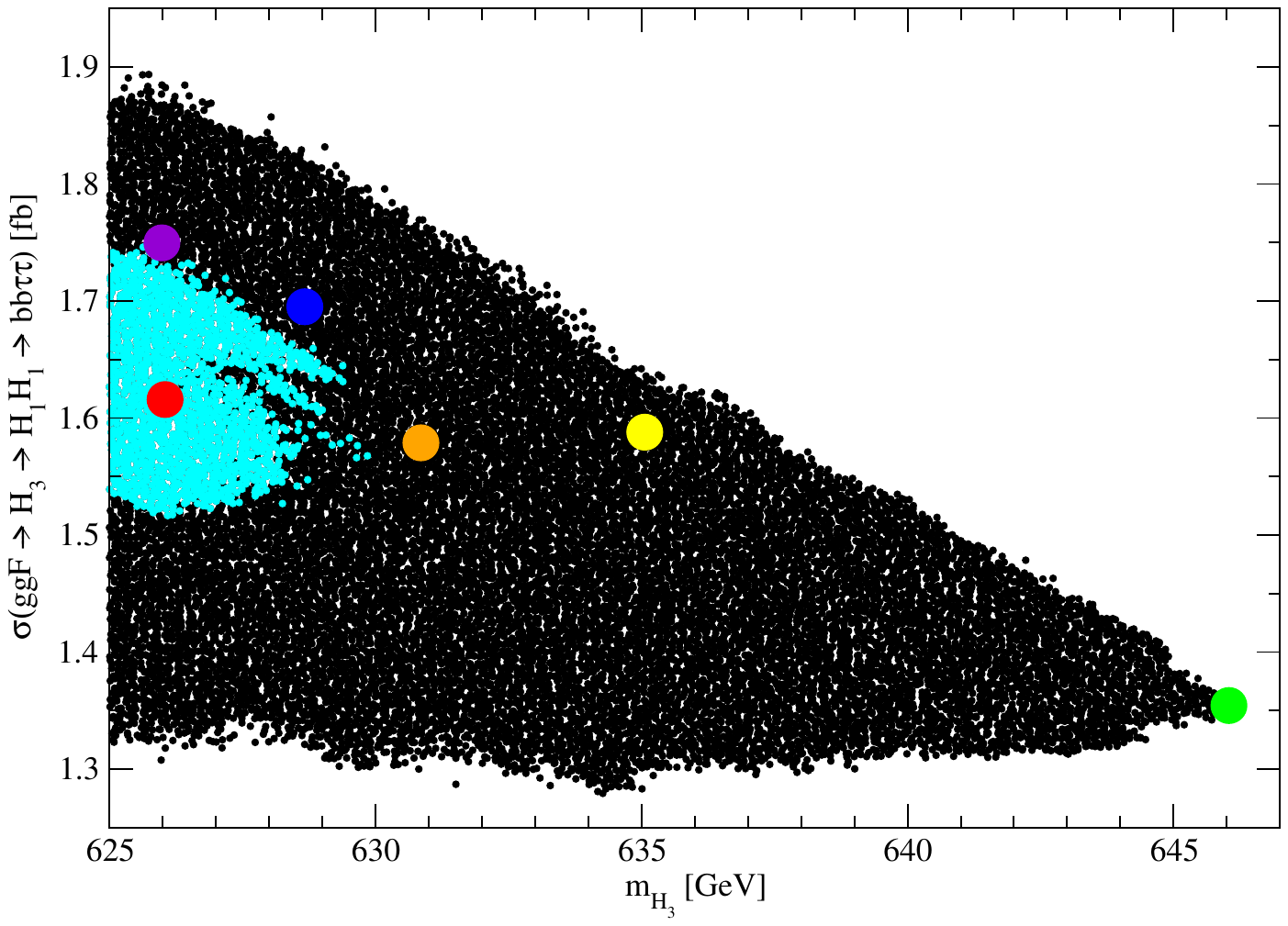} \\
\includegraphics[scale=0.32]{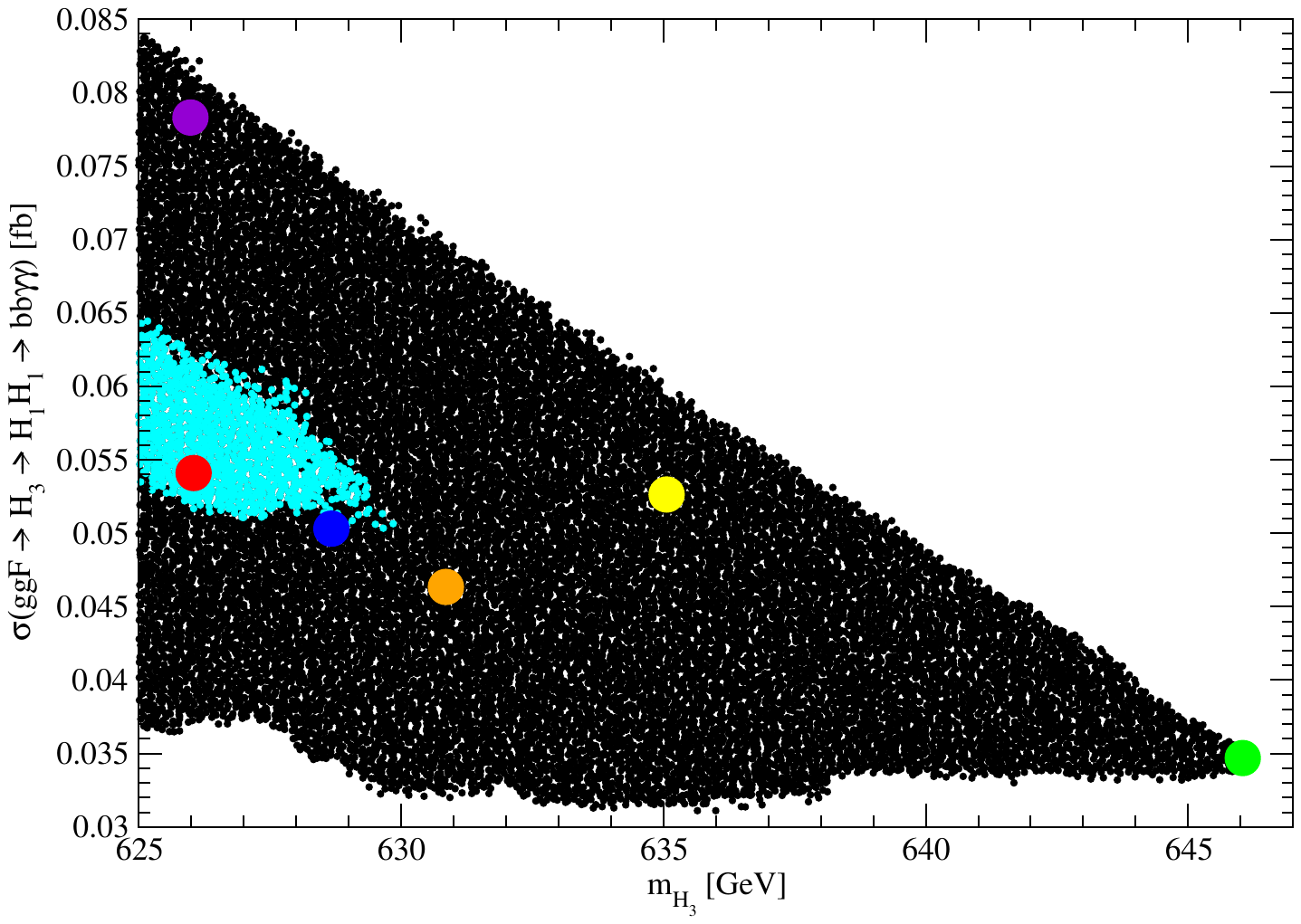}
   & % \hskip -0.5cm
\includegraphics[scale=0.32]{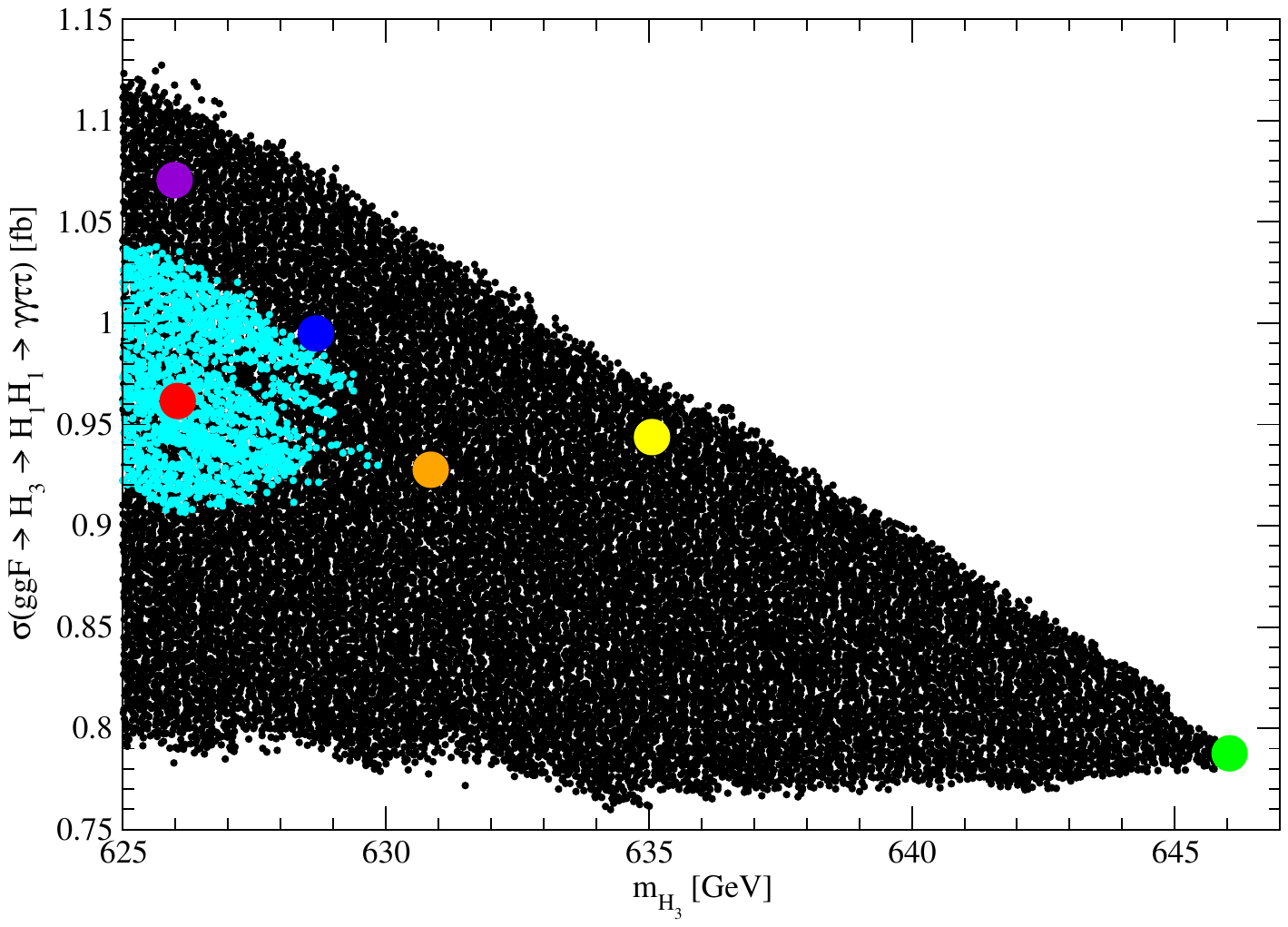}
\end{tabular}
\end{center}
\vskip -1.cm
\caption{Upper left: $\sigma(ggF\to H_3 \to H_1 H_1 \to b\bar{b}b\bar{b}$), upper right: $\sigma(ggF\to H_3 \to H_1 H_1 \to b\bar{b}\tau^+\tau^-$),
lower left: $\sigma(ggF\to H_3 \to H_1 H_1 \to b\bar{b}\gamma\gamma$), lower right: $\sigma(ggF\to H_3 \to H_1 H_1 \to \gamma\gamma\tau^+\tau^-$), all cross sections as a function of the heavy scalar boson mass $M_{H_3}$}
%To check: $b\bar{b}\sim 10\times \tau^+\tau^-$, $\tau^+\tau^- \sim 30\times \gamma\gamma$ ???
\label{fig:4}
\end{figure}

Also the heavy pseudoscalar $A_2$ with its mass close to $M_{H_3}$ gives rise to interesting signatures. Searches for $ggF\to A_2\to Z+H_{SM}$ and for $ggF\to A_2\to Z+H_1$ have been performed by CMS in \cite{CMS:2019qcx,CMS:2019ogx} and by ATLAS in \cite{ATLAS:2022enb}. For $A_2$ masses relevant here, upper limits on the cross sections for $ggF\to A_2\to Z+(H_{SM}\to b\bar{b})$ from CMS \cite{CMS:2019qcx} and from ATLAS \cite{ATLAS:2022enb} are $\sim 30$\,~fb, upper limits on cross sections for $ggF\to A_2\to (Z\to \ell\ell)+(H_1 \to b\bar{b})$ from CMS \cite{CMS:2019ogx} are $\sim 20$\,~fb.
In Figs.~5 we show, both as function of $M_{A_2}$, $\sigma(ggF\to A_2 \to Z+(H_{SM} \to b\bar{b}))$ on the left, and $\sigma(ggF\to A_2 \to (Z\to \ell\ell)+(H_1 \to b\bar{b}))$ on the right. Both cross sections are factors of 20 (for $Z+H_{SM}$) or 5 (for $Z+H_1$) below the limits from ATLAS/CMS, but since the limits from CMS are based on $35.9$\,~fb$^{-1}$ of integrated luminosity the cross section from Fig.~5 are not out of reach in the future. Note that once one multiplies the cross sections into $H_1$ on the right hand side by $1/0.0673$ in order to compensate the BR$(Z\to \ell\ell)$, one finds that these are by a factor~$\sim 30-40$ larger than the cross sections into $H_{SM}$ on the left and side.

%It is remarquable that the cross sections into $H_1$ are nearly twice as large as the cross sections into $H_{SM}$.

\begin{figure}[ht!]
\begin{center}
\hspace*{-10mm}
\begin{tabular}{cc}
\includegraphics[scale=0.32]{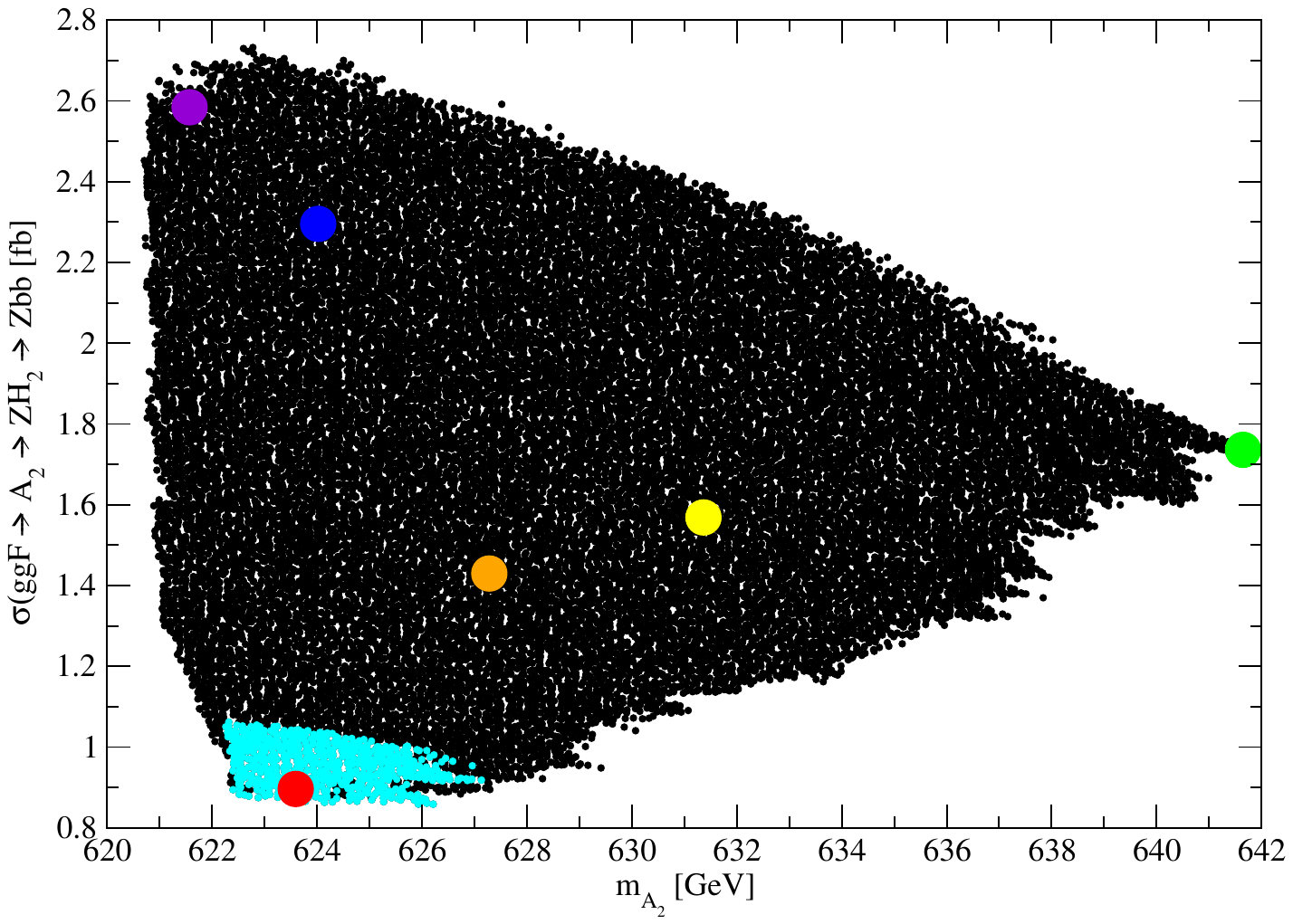}
   & % \hskip -0.5cm
\includegraphics[scale=0.32]{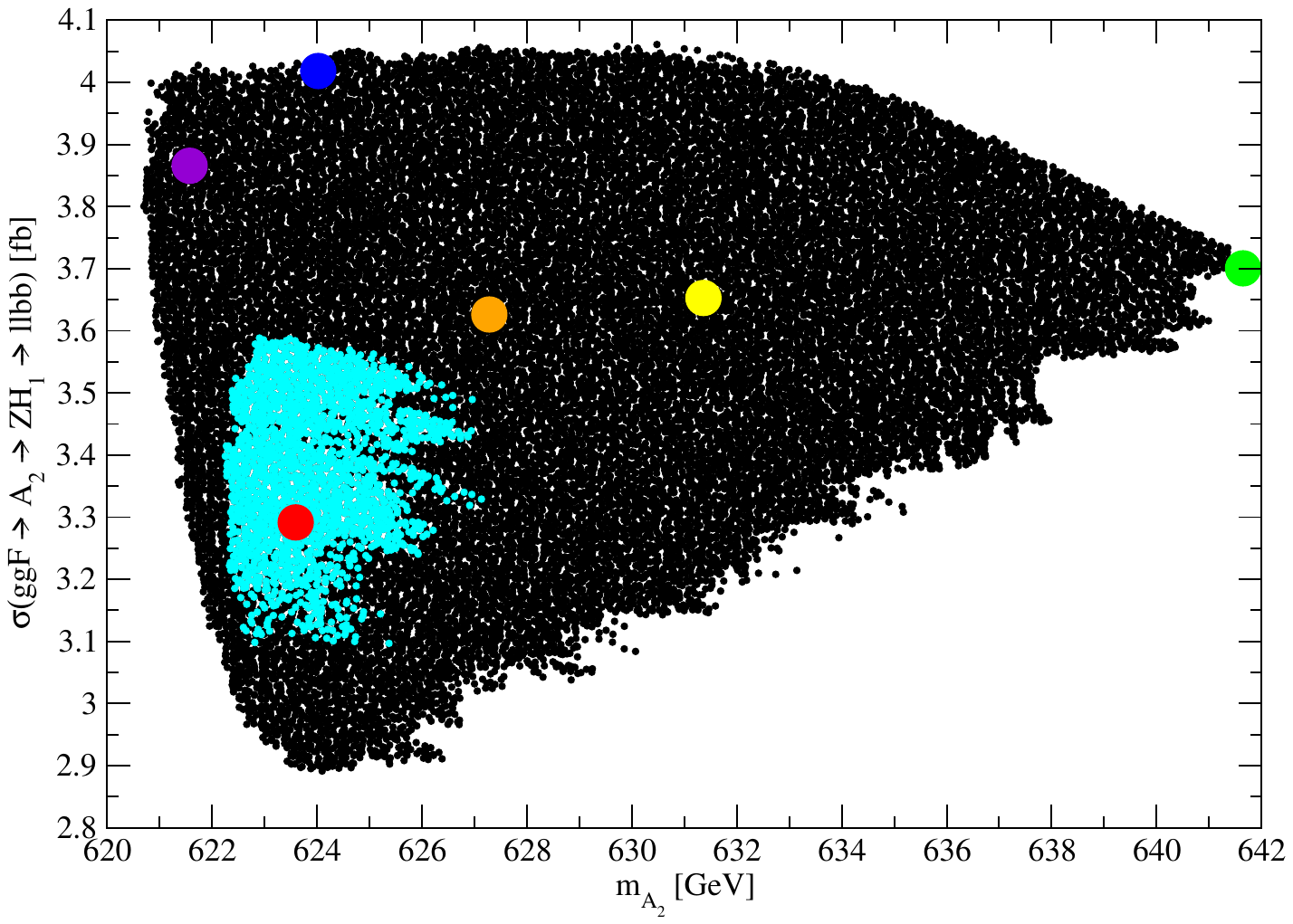}
\end{tabular}
\end{center}
\vskip -1.cm
\caption{Left: $\sigma(ggF\to A_2 \to Z+(H_{SM} \to b\bar{b}))$, right: $\sigma(ggF\to A_2 \to (Z\to \ell\ell)+(H_1 \to b\bar{b}))$, both as function of $M_{A_2}$.}
\label{fig:5}
\end{figure}

{Finally the charged Higgs boson with its mass similar to $M_{H_3}$ might be observable via its dominant decay channel $H^\pm \to t+b$. Recent searches in this channel have been performed by CMS in \cite{CMS:2020imj} (based on 35.9~fb$^{-1}$) and by ATLAS in \cite{ATLAS:2021upq} (based on 139~fb$^{-1}$). For $M_{H^\pm} \sim 600-650$~GeV, the upper limit on $\sigma(pp\to tbH^\pm)\times Br(H^\pm \to tb)$ obtained in \cite{ATLAS:2021upq} is of the order of 150~fb. We have computed the charged Higgs production cross section using results from the 
LHC Higgs Cross Section Working Group \cite{LHCyellowreport} (CERN Yellow Report~4 2016)
based on results in \cite{Flechl:2014wfa,Degrande:2015vpa,LHCHiggsCrossSectionWorkingGroup:2016ypw,Dittmaier:2009np,Berger:2003sm}.
In Fig.~6 we show $\sigma(pp\to tbH^\pm)\times Br(H^\pm \to tb)$ as function of $M_{H^\pm}$; we see that the possible values in the NMSSM scenario presented here are still below the present sensitivities. Actually the branching ratio $Br(H^\pm \to W^\pm + H_1)$ is in the $10-20\%$ range. A search for $\sigma(pp\to (H^\pm\to W^\pm + H))$ has been carried out by CMS in \cite{CMS:2022jqc} assuming, however, $M_H = 200$~GeV and not 95~GeV as it would be the case here.}

\begin{figure}[ht!]
\begin{center}
\hspace*{-10mm}
\begin{tabular}{cc}
\includegraphics[scale=0.4]{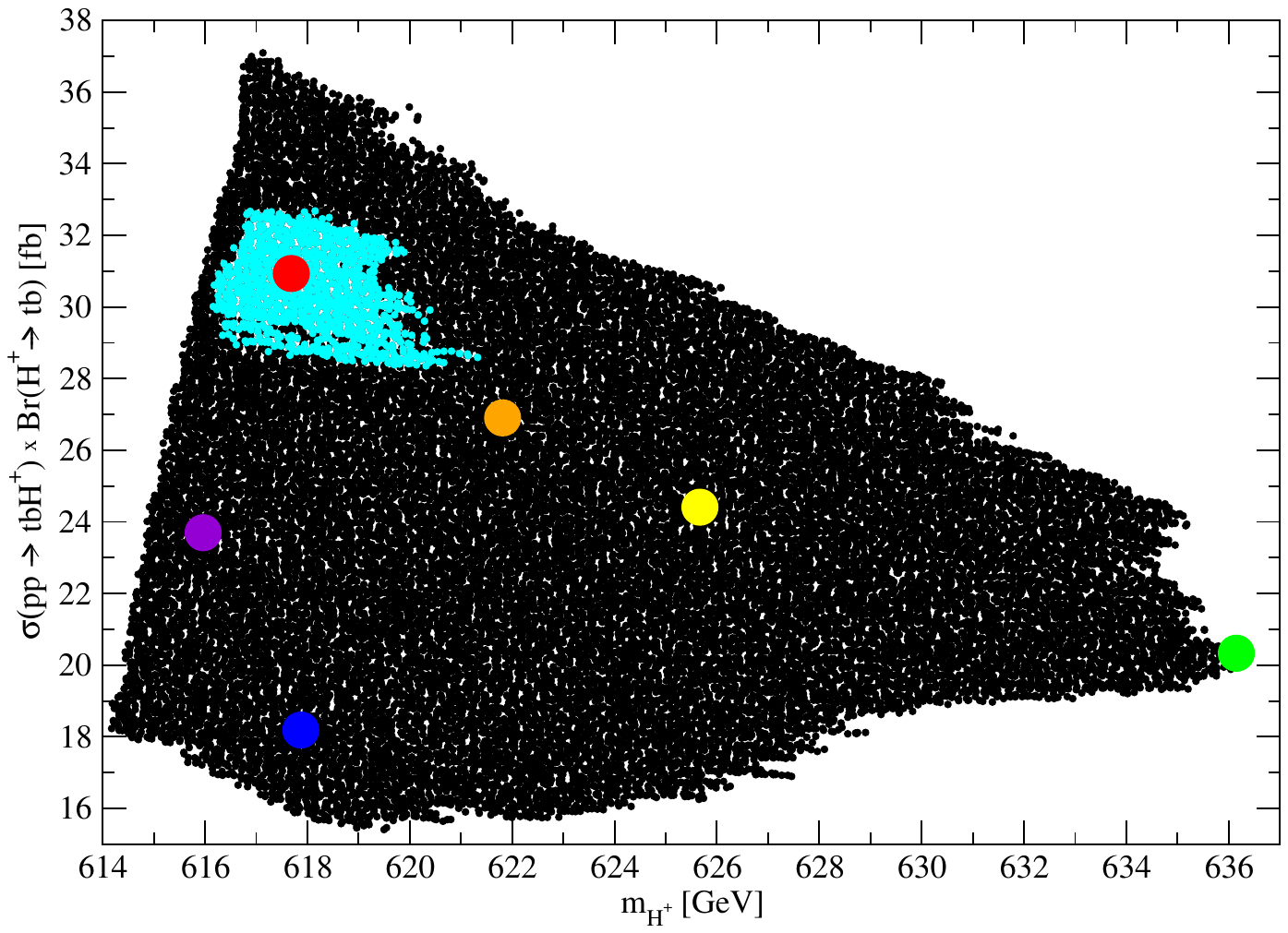}
\end{tabular}
\end{center}
\vskip -1.cm
\caption{{ $\sigma(pp\to tbH^\pm)\times Br(H^\pm \to tb)$ as function of $M_{H^\pm}$.}}
\label{fig:6}
\end{figure}

We end this Section with Tables giving details of some benchmark points satisfying all of the imposed constraints. The benchmark points are chosen such that they cover various regions visible in Figs.$\ 1-5$. NMSSM specific input parameters, $M_{H_3}$, {stop masses and $A_t$} are given in Table~2. {The not too large values for the stop masses and $A_t$ indicate that, compared to the MSSM, less contributions from stop loops are necessary in order to lift the SM Higgs mass to $125.2\pm 3$~GeV.}

\begin{table}[ht]
\begin{center}
\begin{tabular}{| c | c | c | c | c | c | c | c | c |c | c | c |  }
\hline
& $\lambda$  & $\kappa$  & $A_\lambda$ &$A_\kappa$ & $\mu_{\rm eff}$ & $\tan\beta$ & 
$M_{H_3}$ & $M_{t_1}$& $M_{t_2}$ & $A_t$ & colour
\\
\hline
BP1 & 0.670 & 0.327 & 442 & $-495$ & 282 &  2.04 & 626 & 1007 & 1175 & $-508$ & red  \\
\hline
BP2 & 0.657 & 0.388 & 440 & $-552$ & 265 &  2.36 & 646 & 2865 & 2921 & $-215$ & green  \\
\hline
BP3 & 0.661 & 0.376 & 443 & $-489$ & 245 &  2.50 & 629 & 2563 & 2948 & $-152$ & blue  \\
\hline
BP4 & 0.663 & 0.361 & 446 & $-526$ & 269 &  2.24 & 635 & 1090 & 2947 & $-986$ & yellow  \\
\hline
BP5 & 0.653 & 0.378 & 424 & $-537$ & 258 &  2.34 & 626 & 2539 & 2663 & $-197$ & violet  \\
\hline
BP6 & 0.656 & 0.355 & 442 & $-536$ & 272 &  2.18 & 631 & 1172 & 2881 & $-1016$ & orange  \\
\hline
\end{tabular}
\end{center}
\caption{NMSSM specific input parameters, $M_{H_3}$, stop masses and $A_t$ for six benchmark points. }
\label{tab:2}
\end{table}

In Table~3 we show the corresponding (reduced) cross sections as defined in \eqref{mulep}, \eqref{mugamgam} and \eqref{650GeV}, reduced couplings of $H_3$ and $A_2$ to top quarks, as well as cross sections times branching fractions for additional processes involving $H_3$ or $A_2$. The ratio of the cross sections $\sigma^{A_2}_{Z H_1}$ to $\sigma^{A_2}_{Z H_{SM}}$ shows more clearly the factor~$\sim 30-40$ in favour of $\sigma^{A_2}_{Z H_1}$.

\begin{table}[h!]
\begin{center}
\begin{tabular}{| c | c | c | c | c | c | c |}
\hline
  & BP1  & {BP2} & BP3   & BP4  & BP5 & BP6  \\
\hline
\phantom{\Big{(}}   $\mu^{LEP}_{bb}$:     & $1.22\times 10^{-2}$ & $1.90\times 10^{-2}$ 
& $2.68\times 10^{-2}$ & $1.79\times 10^{-2}$ & $2.47\times 10^{-2}$ & $1.49\times 10^{-2}$ \\

$\mu^{LHC}_{\gamma\gamma}$:               & $8.02\times 10^{-2}$ & $8.06\times 10^{-2}$ 
& $1.21\times 10^{-1}$ & $9.96\times 10^{-2}$ & $1.65\times 10^{-1}$& $8.01\times 10^{-2}$ \\

$\sigma_{bb\gamma\gamma}$:                & $9.16\times 10^{-2}$ & $9.01\times 10^{-2}$ 
& $9.01\times 10^{-2}$ & $9.13\times 10^{-2}$ & $9.00\times 10^{-2}$ & $9.81\times 10^{-2}$ \\

$g_{H_3 tt}$:                             & $-0.491$ & $-0.429$ & $-0.404$ & $-0.445$ & $-0.432$ & $-0.630$ \\

$g_{A_2 tt}$:                             & $0.487$ & $0.421$ & $0.398$ & $0.444$ & $0.429$ & $0.456$ \\

\phantom{\Big{[}} $\sigma^{H_1 H_{SM}}_{bb\tau\tau}$:       & $2.65$ & $2.67$& $2.76$ & $2.67$ & $2.72$ & $2.78$ \\

\phantom{\Big{[}} $\sigma^{H_1 H_1}_{bb bb}$:               & $7.09$ & $6.04$& $7.51$ & 6.95 & 7.44 & 7.00   \\

\phantom{\Big{[}} $\sigma^{A_2}_{Z H_{SM}}$:                & $1.51$ & $3.01$& $3.94$ & $2.73$ & $4.45$ & $2.51$ \\

\phantom{\Big{[}} $\sigma^{A_2}_{Z H_1}$:                 & $48.9$ & $55.0$& $59.7$ & 54.3 & 57.4 & 53.9  \\

\phantom{\Big{[}} $\sigma^{H^\pm}_{tb}$:                 & $30.9$ & $20.3$& $18.2$ & 24.4 & 23.7 & 26.9  \\

\hline
\end{tabular}
\caption{$\mu^{LEP}_{bb}$, $\mu^{LHC}_{\gamma\gamma}$ and $\sigma_{bb\gamma\gamma}$ from Section~1, reduced couplings to $tt$ and additional cross sections times branching fractions for processes for the six benchmark points.\\ 
\vspace{.5em}\noindent $\sigma^{H_1 H_{SM}}_{bb\tau\tau}$ denotes the cross section for $ggF\to H_3\to (H_1\to b\bar{b}) + (H_{SM}\to \tau\tau)$,\\ 
\vspace{.5em}\noindent $\sigma^{H_1 H_1}_{bb bb}$ the cross section for $ggF\to H_3\to (H_1\to b\bar{b})+(H_1\to b\bar{b})$,\\ 
\vspace{.5em}\noindent $\sigma^{A_2}_{Z H_{SM}}$ the cross section for $ggF \to A_2\to Z+H_{SM}$, and\\ 
\vspace{.5em}\noindent $\sigma^{A_2}_{Z H_1}$ the cross section for $ggF \to A_2\to Z+(H_{1}\to b\bar{b})$, and\\
\vspace{.5em}\noindent {$\sigma^{H^\pm}_{tb}$ the cross section for $pp \to H^\pm \to tb$.} All cross sections are given in fb.
}
\end{center}
\label{tab:3}
\end{table}

\section{Summary and Conclusions}

In the present paper we have shown which sparticle spectra in the NMSSM can simultaneously describe an extra Higgs boson near 95~GeV, and an excess in the resonant production of SM plus BSM Higgs bosons in the diphoton plus $b\bar{b}$ channel by CMS in \cite{CMS:2023boe} for a heavy resonance of a mass near $\sim 650$~GeV. This region of the parameter space of the NMSSM is limited, amongst others, by a search for $X \to (H_{SM}\to \tau\tau) + (H_1 \to b\bar{b})$ by CMS in \cite{CMS:2021yci} for $M_X=600, 700$~GeV. Still, we find viable regions in the parameter space at the $2\, \sigma$ level. Admittedly this is perhaps not the strongest hint for new physics at present, but we find it worthwhile to underline that this region exists even in the light of the latest results from the LHC, notably in light of the measurements of CMS \cite{CMS:2022dwd} and ATLAS \cite{ATLAS:2022vkf} of the couplings of the SM Higgs boson.

One interesting feature is that relatively light higgsino-like charginos with masses below $\sim 400$~GeV can help to enhance the $BR(H_1 \to \gamma\gamma)$ via loops to the level required by eq.\eqref{mugamgam}, at least within the $2\,\sigma$ level. This also implies relatively light neutralinos, which is visible in the form of $\mu_{\text{eff}}$ for the benchmark points shown.

In the NMSSM, the spectrum of additional Higgs bosons near 650~GeV is necessarily MSSM-like, i.e. consists in nearly degenerate scalars, pseudo-scalars and charged scalars. However, their branching fractions into standard search channels are reduced by their decays into light Higgs bosons, higgsinos and charginos. Searches for the $b\bar{b}$ final state are disfavoured by the low value of $\tan\beta$, searches for the $t\bar{t}$ channel are more promising.
{Likewise, searches for the charged Higgs in the $t \bar{b} + c.c.$ channel can be promising, although the corresponding branching fraction can be somewhat reduced by the decays $H^+\to W^+ + H_S$.}

Cross sections times branching fractions for the production of the MSSM-like sector are shown in Figs.$\ 2-6$, which should help to verify or exclude the NMSSM scenarios presented here in the future. Some of the available searches by ATLAS and CMS already touch the parameter space of the NMSSM, and our tables allow to estimate which future searches can be promising not only using available data, but also after the upgrade of the LHC to High Luminosity after a suitable rescaling. The parameters shown in Tab.$\ 1$ help to clarify which range of NMSSM parameters correspond to these scenarios. It is remarkable that the relevant ranges of large $\lambda$ and small $\tan\beta$ coincide with the ones where a NMSSM-specific uplift of the SM Higgs mass at tree level helps to explain its value well above $M_Z$ \cite{Maniatis:2009re,Ellwanger:2009dp}.

\section*{Acknowledgements}

U.E. acknowledges motivating and helpful discussions with members of the LHC-HXSWG3-NMSSM working group.

\end{document}